\documentclass{article}
\usepackage{subcaption}
\usepackage{arxiv}
\usepackage{array}
\usepackage{amsmath}
\usepackage{tikz}
\usepackage[utf8]{inputenc} 
\usepackage[T1]{fontenc}    
\usepackage{hyperref}       
\usepackage{url}            
\usepackage{booktabs}       
\usepackage{amsfonts}       
\usepackage{nicefrac}       
\usepackage{microtype}      
\usepackage{lipsum}
\usepackage{graphicx}
\usepackage{amsmath}
\usepackage{amsfonts}
\usepackage{algorithm}
\usepackage{algorithmic}
\usepackage[numbers]{natbib}

\DeclareMathOperator*{\argmax}{arg\,max}
\graphicspath{ {./images/} }

\title{Deep multi-intentional inverse reinforcement learning for cognitive multi-function radar inverse cognition}

\author{
 Hancong Feng \\
 School of Information and Communication Engineering\\
 UEST of China\\
  \texttt{202211012303@std.uestc.edu.cn} \\
   \And
 KaiLI Jiang \\
 School of Information and Communication Engineering\\
 UEST of China\\
  \texttt{jiangkelly@foxmail.com} \\
  \And
 Bin tang \\
 School of Information and Communication Engineering\\
 UEST of China\\
}

\begin{document}
\maketitle
\begin{abstract} 
  In recent years, radar systems have advanced significantly, offering environmental adaptation and multi-task capabilities. These developments pose new challenges for electronic intelligence (Elint) and electronic support measures (ESM), which need to identify and interpret sophisticated radar behaviors. This paper introduces a Deep Multi-Intentional Inverse Reinforcement Learning (DMIIRL) method for the identification and inverse cognition of cognitive multi-function radars (CMFR). Traditional Inverse Reinforcement Learning (IRL) methods primarily target single reward functions, but the complexity of CMFRs necessitates multiple reward functions to fully encapsulate their behavior. To this end, we develop a method that integrates IRL with Expectation-Maximization (EM) to concurrently handle multiple reward functions, offering better trajectory clustering and reward function estimation. Simulation results demonstrate the superiority of the proposed method over baseline approaches.
\end{abstract}

\section{Introduction}
In recent years, with the continuous increase in computing capacity, radar systems are becoming more sophisticated and intelligent. Such systems can adaptively select their waveforms \cite{bluntOverviewRadarWaveform2016,kershawOptimalWaveformSelection1994} based on the observed environment and simultaneously perform multiple tasks such as search and tracking \cite{klemmArrayRadarResource2017}. As a result, passive recognition of such complex systems has become a highly demanding task in the field of electronic intelligence (Elint) and electronic support measures (ESM).

Fully adaptive radar (FAR), or cognitive radar (CR), refers to radar systems that can adjust their transmission and reception strategies based on real-time feedback from the observed environment. This concept dates back to the early 2000s when Simon Haykin first introduced the concept of the perception-action cycle (PAC) from cognitive neural science to radar system design. Generally, a FAR can be divided into an assessment branch and a management branch \cite{klemmArrayRadarResource2017}. The assessment branch is responsible for environmental perception, in which target states are estimated from echoes. For this branch, several processing techniques have been developed. For example, space-time adaptive processing (STAP) helps the radar learn statistics of the interference environment to improve the detection of targets, and multi-hypothesis tracking estimates the current target maneuver state to improve tracking performance. On the other hand, the management branch is responsible for the control of the transmitter, especially for Electronically Steered Array (ESA) antennas \cite{charlish2017cognitive}. For this branch, research mainly focuses on management techniques for waveform adaptation, scheduling, and priority assignment. For example, in \cite{selviReinforcementLearningAdaptable2020,thorntonDeepReinforcementLearning2020} and \cite{pulkkinenTimeBudgetManagement2021}, reinforcement and deep reinforcement learning have been introduced for spectrum sharing and task scheduling. In \cite{irci2010study}, a quality of service resource allocation method (Q-RAM) has been studied to achieve optimal real-time resource allocation in phased array radar systems. \cite{zhangLatentMaximumEntropy2024} has provided a comprehensive overview of such algorithms.

Currently, on the reconnaissance side, researchers mainly focus on the analysis of multi-function radar (MFR) pulse streams. One line of research focuses on obtaining meaningful representations of the MFR pulse sequences. Reference \cite{visnevskiHiddenMarkovModels2005} derives a Hidden Markov Model that converts the Time-Of-Arrival (TOA) sequence into symbols. Reference \cite{shuangReconstructionMultifunctionRadar2014} introduces a technique in bioinformatics to reconstruct multi-function radar search plans. Reference \cite{zhuModelBasedTimeSeries2021} utilized a parametric model to represent MFR pulse sequences. Reference \cite{yuanReconstructionRadarPulse2022} introduced a method for reconstructing radar pulse repetition patterns using semantic coding of intercepted pulse trains, enabling a compact PRI representation. Another line of research focuses on state estimation and work modes recognition. References \cite{wang2008signal} and \cite{visnevskiSyntacticModelingSignal2007} described a stochastic context-free grammar (SCFG) approach for describing the task scheduling of the MFR. Reference \cite{ouNovelApproachRecognition2017} proposed a predictive state representation approach for recognizing and predicting multi-function radar behaviors. References \cite{apfeldIdentificationRadarEmitter2020,xuMethodFunctionalState2021,liWorkModesRecognition2020} utilize recurrent neural networks (RNN) to achieve the recognition of work modes from a stream of pulses. Reference \cite{zhaiFewshotRecognitionMultifunction2022} focuses on work modes recognition under few-shot settings and proposes a coding refined prototypical random walk network to improve classification. However, because MFRs are complex systems driven by external events, it is inadequate to identify the system dynamics by just looking at its output.

Recently, inverse reinforcement learning (IRL) has been applied to infer strategies of cognitive radar. The pioneering work was done in \cite{krishnamurthyIdentifyingCognitiveRadars2020}, in which Professor Vikram Krishnamurthy introduced the revealed preferences framework from microeconomics to reconstruct the radar's utility function. Then in the preceding study \cite{zhangLatentMaximumEntropy2024}, missing and non-optimal observations are considered, and a latent maximum entropy inverse reinforcement learning algorithm is proposed. However, cognitive radars often switch between multiple modes or intentions depending on the environment, mission requirements, or threats, so one utility function is inadequate to fully explain their behavior.To address these limitations, this paper focuses on cognitive radars that are capable of performing multiple tasks, namely the cognitive multi-function radar (CMFR). We employ deep multi-intentional inverse reinforcement learning (DMIIRL) to simultaneously estimate multiple reward functions and cluster the trajectories. DMIIRL uses multiple deep neural networks to parameterize the reward functions and associate each trajectory with its corresponding reward through the expectation-maximization (EM) algorithm. By leveraging DMIIRL, we have addressed the challenge of interpreting radar actions that are driven by diverse objectives, obtaining better predictive accuracy of its future actions. The contributions are as follows:
\begin{itemize}
  \item We extend the maximum likelihood IRL proposed in \cite{babes2011apprenticeship} with deep neural networks and apply it to the inverse cognition of CMFR.
  \item We propose to use deep multi-intentional inverse reinforcement learning (DMIIRL) to address the challenge of CMFR inverse cognition.
  \item The effectiveness and superiority of the proposed method are verified by comparison with existing IRL algorithms under various conditions. 
\end{itemize}
\section{Problem formulation}
\subsection{Cognitive multi-function radar}
Cognitive radar is an advanced radar system designed to enhance its performance by acquiring and utilizing knowledge of its operating environment.

Due to the curse of dimensionality, sensor management of CMFR is divided into multiple hierarchies, where lower levels might handle rapid, fine-grained adjustments (e.g., beam steering, frequency agility), while higher levels manage more strategic decisions (e.g., work mode selection). Reference \cite{charlish2017cognitive} illustrates a framework for cognitive radar that involves five information processing levels. At the signal level, raw radar signals are optimized for signal-to-interference ratios. The measurement level converts these signals into data for tracking. At the object level, measurements are integrated to track targets. The situation level interprets tracked objects and their interactions. Finally, the mission level aligns radar operations with overall objectives.

\begin{figure}[htbp]
  \centering
  \includegraphics{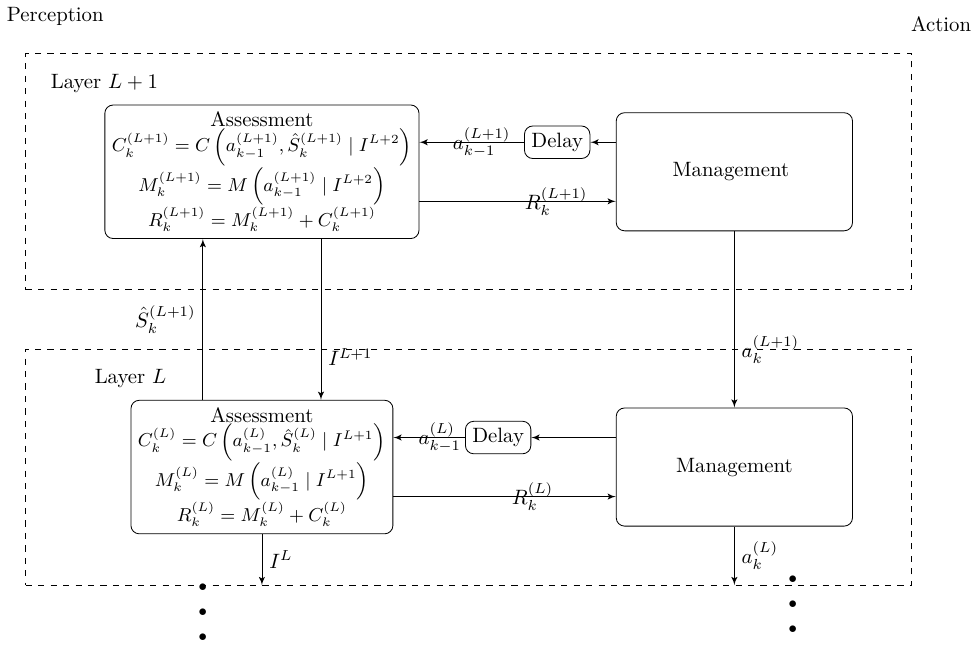}
  \caption{General perception-action cycle for the CMFR with arbitrary layer $L$ and layer $L+1$. The reward function in time step $k$ at layer $L$ is denoted by $R_{k}^{(L)}$, which is equal to the sum of the corresponding performance cost $C_{k}^{(L)}$ and measurement cost $M_{k}^{(L)}$. $I^{L}$ represents the information from layer $L$ that influences lower layers' perceptions. $\hat{S}_k^{(L)}$ is the perception of the environment for layer $L$. For example, at the measurement level, $\hat{S}_k^{(L)}$ can represent the radar tracker’s belief (posterior) of the target maneuver states\cite{krishnamurthyIdentifyingCognitiveRadars2020}.}
  \label{fig:pac}
\end{figure}

Each of these abstraction levels comprises an assessment module for evaluation and a management module for control, forming a perception–action cycle. The assessment module is characterized by its reward (or cost) function, which takes into account the measurement cost and the performance cost \cite{mitchellFullyAdaptiveRadar2018}. The measurement cost is related to resources like radar bandwidth and power used to obtain measurements with specific parameters. The performance cost quantifies the system's effectiveness in performing tasks based on its perception of the environment; for example, the covariance matrix for target tracking. On the other hand, the management module is responsible for making decisions and taking actions that balance the measurement and performance costs. It adjusts the system's sensing parameters (e.g., radar waveform) to optimize the overall performance by balancing the measurement and performance costs. Fig. \ref{fig:pac} depicts two hierarchically connected levels, where information flows bidirectionally within both the perception and action sides of the hierarchy, allowing higher layers to develop more abstract understandings of the environment and influence lower layers' perceptions and actions \cite{mitchellHierarchicalFullyAdaptive2018}.

\begin{figure}[htbp]
  \centering
  \includegraphics{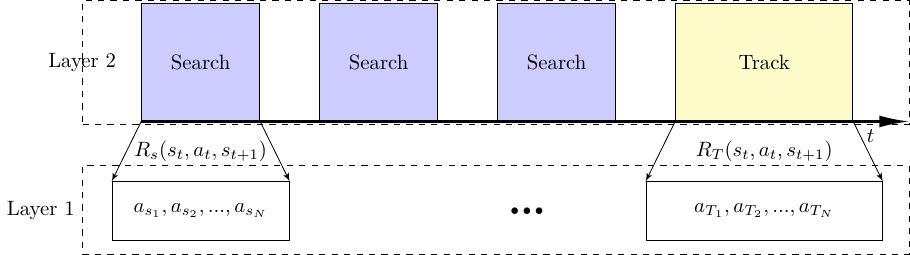}
  \caption{An exemplar radar with two hierarchies. The radar can perform multiple tasks such as search and track. For different tasks, different reward functions are used.  }
  \label{fig:Hierarchy_example}
\end{figure}

Due to the interaction between the higher levels and the lower ones, the goal or preference of the lower PAC can change over time. Take the radar with two sensor management layers as an example (see Fig. \ref{fig:Hierarchy_example}): the first management layer determines the work modes (e.g., search, tracking) of the radar, while the second layer minimizes the costs specified by each work mode. 

\subsection{ESM for CMFR}
Electronic Support Measures (ESM) can be broadly defined as actions taken to analyze intercepted electromagnetic energy for threat recognition and operational planning. For CMFRs, the goal is to mine their signal structure and estimate its rewards, so that we can evaluate the threat and predict future actions.

\begin{figure}[htbp]
  \centering
  \includegraphics{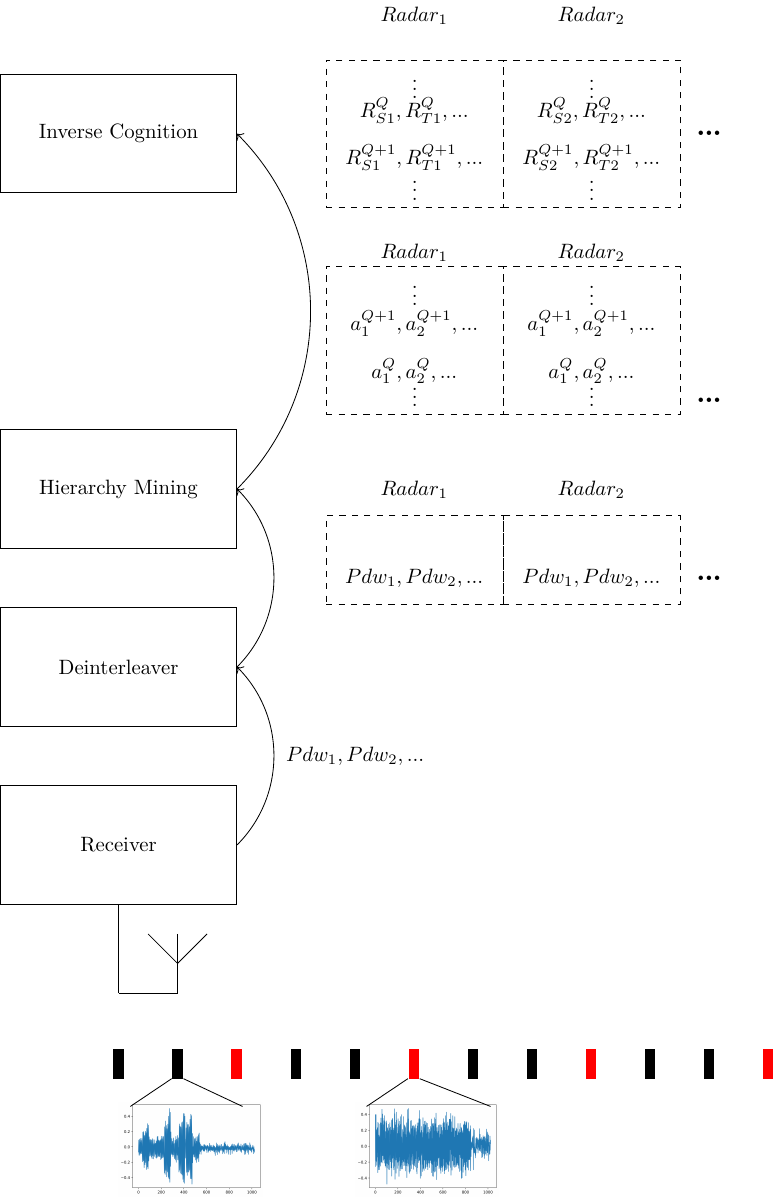}
  \caption{An general ESM framework for CMFRs.}
  \label{fig:esm}
\end{figure}

Fig. \ref{fig:esm} describes an ESM framework for CMFR. It begins with the receiver, which captures and represents the incoming pulse stream as a sequence of Pulse Descriptor Words (PDWs). These PDWs are then processed by the deinterleaver, which sorts them according to their originating radar emitters. The separated PDWs are fed into the hierarchy mining module, where actions at different abstract levels are identified. An action is the basic unit that can be scheduled by the radar resource manager. For example, the early 'Mercury' radar \cite{wangModelingInterpretationMultifunction2008} uses 'radar phrases' as the basic scheduling units, while for more sophisticated radars \cite{apfeldModellingLearningPrediction2019}, smaller units could be used for scheduling. With the extracted actions, the inverse cognition module then estimates the rewards that best explain the actions taken by the CMFR.
The receiver and deinterleaver have been well studied, and algorithms for hierarchy mining have been recently discussed \cite{baoBayesianNonparametricHidden2023,liuRecognitionMultifunctionRadars2020}. Therefore, this study only focused on estimating the CMFR's rewards and replicating its behavior.

Unlike previous research \cite{krishnamurthyIdentifyingCognitiveRadars2020,zhangLatentMaximumEntropy2024}, even actions for the same layer could be generated by models with different rewards because of the control of its preceding layer. Therefore, we may need multiple rewards to fully explain the CMFR's behavior. 

\section{Method}
In this section, we will first introduce some preliminaries about Reinforcement Learning and Inverse Reinforcement Learning in the context of radar inverse cognition. Then, we will introduce the method proposed to simultaneously estimate multiple reward functions and associate trajectories likely generated by them.

\subsection{Reinforcement Learning and Inverse Reinforcement Learning}

Reinforcement Learning (RL) is a learning paradigm where an agent interacts with an environment to learn a policy that maximizes cumulative rewards. Formally, RL is typically modeled as a Markov Decision Process (MDP), which is defined by the tuple $(S, A, P, R, \gamma)$:
\begin{itemize}
    \item $S$ is a finite set of states.
    \item $A$ is a finite set of actions.
    \item $P: S \times A \times S' \rightarrow [0, 1]$ is the state-transition probability function.
    \item $R: S \times A \times S' \rightarrow \mathbb{R}$ is the reward function.
    \item $\gamma \in (0, 1]$ is the discount factor.
\end{itemize}
Here, the reward function $R$ depends on both.

The value function of a policy $\pi$ is defined as the expected cumulative reward at a given state when following the policy:
\begin{equation}
  V^\pi(s) = \mathbb{E} \left[ \sum_{t=0}^{\infty} \gamma^t R(s_t, a_t, s_{t+1}) \mid s_0 = s, a_t \sim \pi(s_t) \right]
  \label{eq:value_function}
\end{equation}
and the action-value function is defined as:
\begin{equation}
  Q^\pi(s, a) = \mathbb{E} \left[ \sum_{t=0}^{\infty} \gamma^t R(s_t, a_t, s_{t+1}) \mid s_0 = s, a_0 = a, a_t \sim \pi(s_t) \right]
  \label{eq:action_value_function}
\end{equation}
where $t = 0, 1, \ldots, T$ are the discrete time indices. The relationship between $V^\pi(s)$ and $Q^\pi(s, a)$ can be expressed by the Bellman equation:
\begin{equation}
  Q^\pi(s, a) = \mathbb{E} \left[ R(s, a, s') + \gamma V^\pi(s') \mid s' \sim P(s' \mid s, a) \right]
  \label{eq:bellman_action_value}
\end{equation}
where $P(s' \mid s, a)$ is the transition probability from state $s$ to state $s'$ given action $a$.

The objective in RL is to find an optimal policy $\pi^*: S \rightarrow A$ that maximizes the expected cumulative reward as:

\begin{equation}
  \pi^* = \argmax_\pi V^\pi(s)
  \label{eq:optimal_policy}
\end{equation}
Well-studied algorithms exist to solve this problem, including model-based methods \cite{moerland2023model} and model-free methods \cite{ccalicsir2019model}. Some of these approaches have been applied to the design of cognitive radars \cite{thorntonDeepReinforcementLearning2020,selviReinforcementLearningAdaptable2020}. Algorithm \ref{alg:value_iteration} shows a model-based method called value iteration that directly solves the MDP, which we will refer to many times in the later discussion.
\begin{algorithm}
  \caption{Value Iteration Algorithm}
  \label{alg:value_iteration}
  \begin{algorithmic}
  \STATE \textbf{Initialize} $V(s) \leftarrow 0$ for all $s \in S$
  \REPEAT
      \STATE $\Delta \leftarrow 0$
      \FOR{each $s \in S$}
          \STATE $v \leftarrow V(s)$
          \STATE $V(s) \leftarrow \max_{a \in A} \sum_{s' \in S} P(s' \mid s, a) \left[ R(s, a, s') + \gamma V(s') \right]$
          \STATE $\Delta \leftarrow \max(\Delta, |v - V(s)|)$
      \ENDFOR
  \UNTIL{$\Delta < \epsilon$} \COMMENT{where $\epsilon$ is a small positive number for convergence}
  \end{algorithmic}
\end{algorithm}
On the other hand, Inverse Reinforcement Learning (IRL) is the process of inferring the reward function $R$ given observed behavior from an expert. Formally, given a set of expert trajectories $\mathcal{D} = \{\tau_1, \tau_2, \ldots, \tau_n\}$, where each trajectory $\tau$ is a sequence of state-action pairs $(s_1, a_1), (s_2, a_2), \ldots, (s_T, a_T)$ and the MDP without $R$, IRL aims to solve:
\begin{equation}
  \hat{R} = \arg \max_R \sum_{\tau \in \mathcal{D}} \log P(\tau \mid R)
  \label{eq:estimated_reward}
\end{equation}
  
Here, $P(\tau \mid R)$ denotes the likelihood of trajectory $\tau$ given the reward function $R$. This is an underdetermined problem where multiple reward functions can lead to the same behavior. Therefore, additional assumptions, such as maximum margin \cite{ng2000algorithms} and maximum entropy \cite{ziebart2008maximum}, need to be introduced to obtain a unique solution.

For ESM, we cannot get the beliefs of the radar; thus, estimations must be performed. This can be summarized as:
\begin{itemize}
  \item True environmental dynamics: $s_t \sim P_s(s \mid s_{t-1}, a_{t-1}), s_0 \sim \pi_0$;
  \item Radar observations: $s^{r}_{t} \sim P_{s^{r}_{t}}(s \mid s_t, a_{t-1});$
  \item ESM environmental observations: $s^{e}_{t} \sim P_{s^e_{t}}(s \mid s_t)$;
  
  \item Environmental dynamics estimated by radar: $s^{r}_t \sim P_s^{r}(s \mid s_{t-1}, a_{t-1})$, $s^{r}_0 \sim \pi^{r}_0$;

  \item Environmental dynamics estimated by ESM: $s^e_t \sim P_s^e(s \mid s_{t-1}, a_{t-1}), s^e_0 \sim \pi^e_0$;

  \item Radar actions: $a_t \sim P_a(a \mid s^r_{t})$,
  \item ESM estimation of radar actions: $\hat{a}_t \sim P_{\hat{a}}(a \mid a_t)$.
\end{itemize}
Here, $\sim p(\cdot)$ represents distributed according to a generic conditional probability density (mass) function $p(\cdot)$. This definition extends the radar-target interaction proposed in \cite{krishnamurthyIdentifyingCognitiveRadars2020}, where the ESM receiver also needs to estimate the environmental dynamics.

For ESM, this information asymmetry leads to non-ideal observations in both environmental dynamics and the radar actions, which has been partly studied in \cite{zhangLatentMaximumEntropy2024}.

Because this study mainly focuses on the inverse cognition of CMFR, to simplify the discussion, we do not perform extra data processing for such non-ideal situations and leave it for future research.

\subsection{Deep Maximum Entropy and Maximum Likelihood IRL for cognitive radar reward function estimation}
In this section, we will explain the principle of Maximum Entropy and Maximum Likelihood IRL and how to parameterize an arbitrary reward function of a cognitive radar with neural networks.

The principle of Maximum Entropy IRL was first introduced by Ziebart et al. \cite{ziebart2008maximum}. It assumes that the expert's behavior is not only rational but also maximally random given the constraints of matching feature expectations.

\begin{figure}[htbp]
  \centering
  \includegraphics{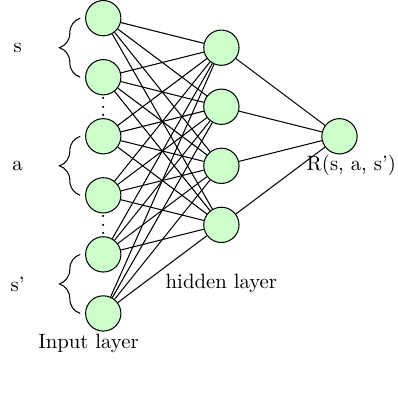}
  \caption{The neural network for reward function parameterization. The input is the concatenation current state $s$, current state action $a$ and the next state $s'$ }
  \label{fig:nn}
\end{figure}

Ziebart et al. have shown that the probability of a trajectory \( \tau \) under the maximum entropy principle is approximately given by:
\begin{equation}
  P(\tau \mid R) = \frac{1}{Z(R)} \exp \left( \sum_{t=1}^{T} R(s_t, a_t,s_{t+1}) \right)
  \label{eq:probability_MaxEnt}
\end{equation}
  
where \( Z(R) \) is the partition function defined as:
\begin{equation}
  Z(R) = \sum_{\tau} \exp \left( \sum_{t=1}^{T} R(s_t, a_t,s_{t+1}) \right).
  \label{eq:partition_function}
\end{equation}
If we parameterize the reward with a neural network (shown in Fig. \ref{fig:nn}) with the tuple $(s_t, a_t, s_{t+1})$ as input, the log-likelihood in \eqref{eq:estimated_reward} can be written as:
\begin{equation}
  L(\theta)= \sum_{\tau \in \mathcal{D}} \log \left(\frac{1}{Z(R)} \exp \left( \sum_{t=1}^{T} R(s_t, a_t,s_{t+1}\mid \theta) \right)\right),
  \label{eq:log-likelihood_maxent}
\end{equation}
where $\theta$ are the trainable parameters of the network.

Substituting the transition probability into \eqref{eq:log-likelihood_maxent} and performing some simplification, the gradient of the log-likelihood with respect to $\theta$ can be written as:
\begin{equation}
  \frac{\partial L(\theta)}{\partial \theta}
  = \sum_{s \in S}\sum_{a \in A}\sum_{s' \in S} (D_{obs}(s,a,s') - D_{\pi}(s,a,s'))\frac{\partial R(s,a,s')}{\partial \theta}.
  \label{eq:gradient_maxent}
\end{equation}
 where $D_{obs}(s,a,s')$ is the state-action-state visitation frequency of the expert estimated from the trajectories dataset, and $D_{\pi}(s,a,s')$ is the expected state-action-state visitation frequency under the policy \( \pi \) derived by the current reward. This can be efficiently computed with dynamic programming \cite{kitani2012activity,ziebart2008maximum}.
 
 With the computed gradient, we can use gradient ascent to maximize the log likelihood, and this can be done automatically with some software frameworks like Pytorch \cite{paszkePyTorchImperativeStyle2019}. The complete procedure is detailed in Algorithm \ref{alg:maxent}.

 \begin{algorithm}
  \caption{Deep Maximum Entropy IRL}
  \label{alg:maxent}
  \begin{algorithmic}[1]
  \STATE \textbf{Input:} $D, S, A, \gamma,N_{iter}$
  \STATE Estimate $D_{obs}(s,a,s')$ and $P$ using expert trajectories $D$
  \STATE $\text{NN} = \text{initialise\_weights}()$
  \FOR{$n = 1 : N_{iter}$}
      \STATE $R(s,a,s') = \text{NN}.forward(s,a,s')$ 
      \STATE \textbf{Solve MDP with current reward}
      \STATE $\pi^n = \text{value\_iteration}(R(s,a,s').detach(), S, A, P, \gamma)$
      \STATE \textbf{Dynamic programming to get the expected state-action-state visitation frequency}
      \STATE $D_{\pi}(s,a,s') = \text{propagate\_policy}(\pi^n, S, A, P)$
      \STATE \textbf{Compute network gradients and update weights}
      \STATE $r\_grads=D_{obs}(s,a,s') - D_{\pi}(s,a,s')$
      \STATE $r\_grads.backward(- r\_grads)$ 
      \STATE optimizer.step()
  \ENDFOR
  \STATE \textbf{Output:} optimal weights $\theta^*$
  \end{algorithmic}
  \end{algorithm}

In addition to MaxEnt IRL, there is a simple and intuitive method that directly maximizes the likelihood of the observed trajectories. This approach was first introduced in \cite{babes2011apprenticeship}, and here we extend this method with a neural network.

First, we find the current Q function with the current estimation of the reward by replacing the 'max' operation in the standard value iteration update with Boltzmann exploration:
\begin{equation}
  Q(s, a\mid \theta) = \sum_{s'} (R(s, a,s'\mid \theta) + \gamma P(s, a, s') \sum_{a} Q(s,a\mid \theta)\pi_{\theta}(s, a)),
  \label{value_iteration_soft}
\end{equation}
and $\pi_{\theta}(s, a)$ is the Boltzmann exploration policy defined as follows:
\begin{equation}
   \pi_{\theta}(s, a)= e^{\beta Q(s,a\mid \theta)} / \sum_{a} e^{\beta Q(s,a\mid \theta)}.
   \label{eq:be_policy}
\end{equation}
In Eq. \ref{eq:be_policy}, $\beta$ is the temperature parameter for Boltzmann exploration. When $\beta=1$, the induced policy is the same as MaxEnt IRL. This allows us to tune the randomness of the learned policy to cope with different levels of noisy demonstrations. With the policy, the maximum likelihood objective is given by:
\begin{equation}
  L(\theta)=\sum_i w_i \sum_{(s, a) \in \tau} \log \left(\pi_{\theta}(s, a)\right),
\end{equation}
where $w_i \in \left\{w_1, \ldots, w_{n}\right\}$ is the occurrence frequency of trajectory $i$ in $\mathcal{D}$ and can be evaluated with the transition probability. Because of the softmax Boltzmann exploration, $L(\theta)$ is differentiable, thus gradient ascent can be used directly for parameter updates. The whole process is summarized in Algorithm \ref{alg:ml-irl}. In practical implementation, the value iteration only runs for a fixed number of times to reduce computational cost.

\begin{algorithm}
  \caption{Deep Maximum Likelihood IRL}\label{alg:ml-irl}
  \begin{algorithmic}[1]
  \STATE \textbf{Input:} $D, S, A, \gamma,\beta,M,N_{iter}$
  \STATE Estimate $P$ using expert trajectories $D$
  \STATE Compute trajectory weights $\left\{w_1, \ldots, w_{n}\right\}$ from $D$ 
  \STATE $\text{NN} = \text{initialise\_weights}()$
  \STATE $Q(s, a)=zeros()$
  \STATE $\pi_{\theta_t}(s,a)=uniform()$
  \FOR{$t = 1$ to $M$}
      \STATE $R(s,a,s') = \text{NN}.forward(s,a,s')$ 
      \FOR{$i = 1$ to $N_{iter}$}
        \STATE $Q(s, a) = \sum_{s'} (R(s,a,s') + \gamma P(s, a, s') \sum_{a} Q(s,a)\pi_{\theta}(s, a))$
      \ENDFOR
        \STATE  $\pi_{\theta_t}(s, a)= \frac{e^{\beta Q(s,a)}}{\sum_{a} e^{\beta Q(s,a)}}$
      \STATE $loss = -\sum_i w_i \sum_{(s,a) \in \xi} \log(\pi_{\theta_t}(s,a))$
      \STATE $loss.backward()$ 
      \STATE optimizer.step()
  \ENDFOR
  \STATE \textbf{Output:} optimal weights $\theta^*$
  \end{algorithmic}
  \end{algorithm}

\subsection{Deep Multiple Intentions IRL For CMFR}
\label{sec:DMLIRL}
So far, we have discussed the fundamentals of Reinforcement Learning and Inverse Reinforcement Learning, particularly the Maximum Entropy and Maximum Likelihood methods with the reward function parameterized by a deep neural network. These methods work well when the expert trajectory dataset \(D\) is generated by a single reward function. However, for CMFR, multiple tasks are performed in a time division multiplexing manner (see Fig.\ref{fig:Hierarchy_example}) and for different tasks, different reward functions are used \cite{mitchellFullyAdaptiveRadar2018}.

We decide to tackle this problem with Multiple Intentions IRL \cite{babes2011apprenticeship}, which simultaneously estimates the reward functions and clusters trajectories according to the tasks. Through this approach, we not only infer the rewards but also identify the tasks.
Formally, given a set of expert trajectories \( D = \{ \tau_1, \tau_2, \ldots, \tau_n \} \), our goal is to estimate a set of reward functions \( \{ R_1, R_2, \ldots, R_K \} \) and associate each trajectory \(\tau_i\) with one of these reward functions. We introduce a latent variable \(z_i \in \{ 1, 2, \ldots, K \} \) that indicates the index of the reward function responsible for generating each trajectory \(\tau_i\). The log-likelihood of the observed trajectories is then given by:
\begin{equation}
  L(\theta) = \sum_{i=1}^N \log \left( \sum_{k=1}^K P(\tau_i \mid R_{\theta_k})\pi_k \right)
  \label{eq:likelihood_mix}
  \end{equation}
  
where \(\theta_k\) represents the parameters of the neural network used to model the \(k\)th reward function, and \(\pi_k=P(z_i = k)\) is the prior probability of selecting reward function \(R_k\).
Similar to the solution to many probability mixture models, we employ an Expectation-Maximization (EM) algorithm to maximize the log-likelihood. The algorithm iterates between the E-step, which computes the posterior probabilities of the latent variables given the current estimates of the reward functions, and the M-step, which updates the network parameters to maximize the lower bound of the log-likelihood.
In the E-step, we compute the expected log-likelihood (EM Q function):
 \begin{equation}
  Q_{em}\left(\Theta \mid \Theta^{(t)}\right)=\mathbb{E}_{Z \mid X, \Theta^{(t)}}[\log p(X, Z \mid \Theta)]=\sum_{i=1}^n \sum_{k=1}^K \gamma_{i k}^{(t)}\left(\log \pi_k+\log P(\tau_i \mid R_{\theta_k})\right),
  \label{eq:Q_em}
\end{equation}
and
\begin{equation}
  \gamma_{i k}^{(t)}=\frac{\pi_k^{(t)} P(\tau_i \mid R_{\theta_k^{(t)}})}{\sum_{j=1}^K \pi_j^{(t)} P(\tau_i \mid R_{\theta_j^{(t)}})}.
  \label{eq:responsibility}
\end{equation}
Where $\Theta=\{\theta_1,\theta_2...\theta_K;\pi_1,\pi_2..\pi_K\}$ denotes all the trainable parameters, and $\gamma_{ik}^{(t)}$ is the probability that trajectory $i$ belongs to reward $\theta_k$ under the parameters of step $t$. In the M-step, we update the reward function parameters together with the mixture weight $\pi_k$ by maximizing the expected log-likelihood:
\begin{equation}
  \Theta^{(t+1)} = \arg \max_{\Theta}Q_{em}\left(\Theta \mid \Theta^{(t)}\right).
  \label{eq:optimization}
\end{equation}
This can be achieved by carrying out the weighted standard IRL independently for each model because $\gamma_{i k}^{(t)}$ are constant. The whole algorithm is summarized in Algorithm \ref{alg:DMIRL}. Like many clustering problems, the number of mixture components may be unknown and has to be estimated from the dataset. So in practical implementation, we set $K$ to a large number and merge the reward functions that result in similar behaviors.

With the estimated $\Theta$, for a newly arrived trajectory $\tau_i$, we can compute the policy with Eq. \eqref{eq:be_policy}. With the policy, the trajectory probability $P(\tau_i \mid R_{\theta_k^{(t)}})$ can be found. By substituting into Eq. \eqref{eq:responsibility} and selecting the model with the highest probability, we can find the reward function that is responsible for generating $\tau_i$ as:
\begin{equation}
  z_i = \arg \max_{k}\gamma_{i k}.
  \label{eq:assignment}
\end{equation}
Through this approach, we not only approximate the rewards but also associate each trajectory of the CMFR with a specific task, providing a more complete description of its behavior.
\begin{algorithm}
  \caption{Deep Multiple Intentions IRL}
  \label{alg:DMIRL}
\begin{algorithmic}[1]
  \STATE \textbf{Input:} $D, S, A, K, \gamma,N_{iter}$
  \STATE Initialize neural network parameters $\theta_1,\theta_2,...,\theta_K$ 
  \STATE Initialize prior probabilities $\theta_K;\pi_1,\pi_2..\pi_K$.
  \FOR{$t = 1$ to $N_{iter}$}
  \STATE \textbf{E-step:} Compute posterior probabilities
  \FOR{each trajectory $\tau_i \in D$}
  \FOR{each reward function $R_k$}
  \STATE Compute $\gamma_{i k}$ using \eqref{eq:responsibility}.
  \ENDFOR
  \ENDFOR
  \STATE \textbf{M-step:} Update neural network parameters and the prior probabilities.
  \FOR{$k = 1$ to $K$} 
  \STATE Using Algorithm\ref{alg:maxent} or Algorithm\ref{alg:ml-irl} to update $\theta_k$ with $\gamma_{i k}$ as the weight.
  \STATE $\pi_k=\frac{1}{n} \sum_{i=1}^n \gamma_{i k}$  
  \ENDFOR 
  \ENDFOR
  \STATE \textbf{Output:} Estimated $\theta_1,\theta_2,...,\theta_K$ and prior probabilities $\theta_K;\pi_1,\pi_2..\pi_K$.
  \end{algorithmic}
  \end{algorithm}
 \section{SIMULATION}
 In this section, a hypothetical CMFR is designed to test the proposed method on trajectory clustering and reward function estimation.  
 \subsection{Simulation Settings}
\begin{figure}[htbp]
  \centering
  \includegraphics[width=0.7\textwidth]{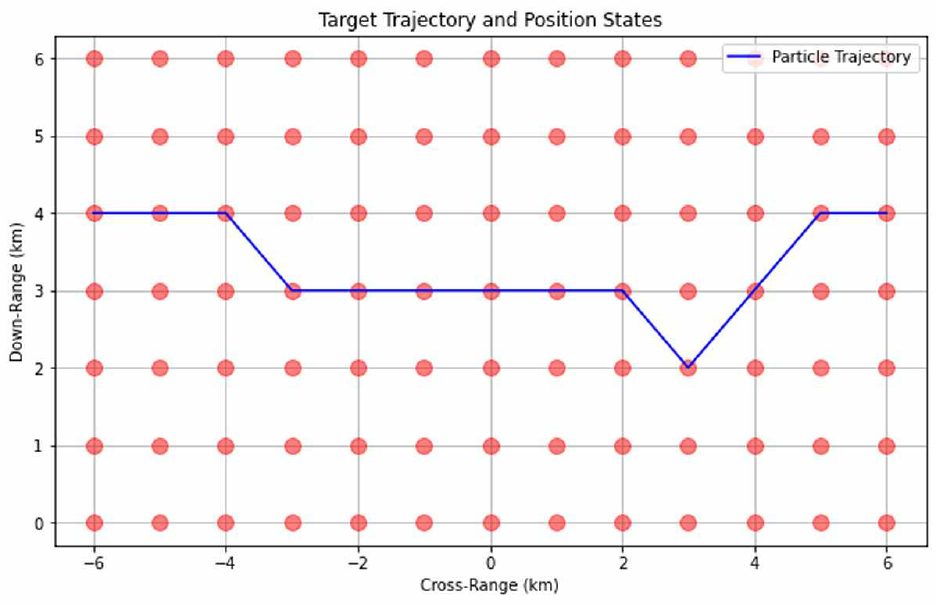}
  \caption{Simulation scene and an exemplar trajectory of the point target.}
  \label{fig:trajectory}
\end{figure}

 IRL is extremely computationally intensive, especially when the state and action space is large. To reduce the simulation time, we adopt a modified version of a radar communication coexistence scenario that is introduced in \cite{selviReinforcementLearningAdaptable2020, thorntonDeepReinforcementLearning2020}. In this environment, a monostatic radar with a Linear Frequency Modulated (LFM) waveform tries to measure a single point target and avoid the interference of a jammer over numerous time steps. The position states $\mathcal{X}$ and velocity states $\mathcal{V}$ are defined as sets of 3D vectors, with $\mathcal{X} = \{\mathbf{x}_1, \mathbf{x}_2, \ldots, \mathbf{x}_\rho\}$ and $\mathcal{V} = \{\mathbf{v}_1, \mathbf{v}_2, \ldots, \mathbf{v}_v\}$, where $\mathbf{x}_i = [x, y, z]$ and $\mathbf{v}_i = [v_x, v_y, v_z]$ represent position and velocity components, respectively. The radar is located at the origin $(0,0,0)$, and Fig \ref{fig:trajectory} shows the trajectory viewed from above. The radar and the jammer are sharing a 100 MHz bandwidth, which is divided into five equally sized sub-channels. The usage of the channel is represented with a length-5 binary vector as $\boldsymbol{\alpha}=\left[\alpha_1, \alpha_2, \ldots, \alpha_5\right], \alpha_i \in \{0,1\}$. For example, $\boldsymbol{\alpha}=\left[1, 0, 0, 1, 0\right]$ means the first and fourth channels have been used. Thus, the state space of the MDP is defined as $S=[\mathcal{X}, \mathcal{V}, \boldsymbol{\alpha}]$, which contains all the combinations of target position states, target velocity states, and interference states. Like previous research \cite{zhangLatentMaximumEntropy2024, thorntonDeepReinforcementLearning2020}, quantization is performed on $\mathbf{x}_i$ and $\mathbf{v}_i$ to maintain tractability.

 Actions of the radar include the radar's selected transmission bands and transmit power, represented by the tuple $(\alpha_i, pow_i)$. Valid actions use only contiguous bands because the radar uses LFM waveform. For example, $(\left[0,0,1,1,0\right], 5)$ is a valid action, while $(\left[1,0,0,1,0\right], 5)$ is not. The additional control of transmit power provides more flexibility in designing reward functions. Other important parameters are summarized in Table \ref{tab:main_para}.
 \begin{table}[h!]
  \centering
  \caption{Main Parameters for the simulation scene.}
  \label{tab:main_para}
  \begin{tabular}{|>{\raggedright\arraybackslash}m{10cm}|>{\centering\arraybackslash}m{4cm}|}
  \hline
  \textbf{Parameter} & \textbf{Value} \\ \hline
  Coherent Processing Interval SNR at Maximum Range & 3 dB \\ \hline
  Pulse Repetition Frequency & 200 $\mu$s \\ \hline
  Waveform & Linear Frequency Modulation \\ \hline
  Pulse Width & 10 $\mu$s \\ \hline
  Maximum Bandwidth & 100 MHz \\ \hline
  Radar Cross Section & $0.1 \, \text{m}^2$ \\ \hline
  Interference-to-Noise Ratio (INR) at the radar & 14 dB \\ \hline
  Coherent processing interval (CPI) & 20 pulses \\ \hline
  Transmit Power & 10 levels from 11 $dbm$ to 20 $dbm$  \\ \hline
  Maximum trajectory length & 20  \\ \hline
  \end{tabular}
  \label{table:parameters}
  \end{table}
 The reward, determined after each state transition, considers the signal-to-interference-plus-noise ratio (SINR), radar bandwidth usage, and the transmit power:
 \begin{equation}
  R_t = 
  \begin{cases} 
    w_1(\text{SINR}) + w_2(\text{Bandwidth})-w_3(\text{Transmit Power}), & \text{if } \text{SINR} \geq 0 \\
  \text{Large Penalty} & \text{if } \text{SINR} < 0, 
  \end{cases}
\end{equation}
, where $w_1$, $w_2$, $w_3$ are the positive weights that determine the importance of each quantity. In general, we encourage the radar to maximize its detection or tracking performance while lowering the transmission power. Depending on the specific application (e.g., Search, Track) or goals (e.g., energy minimization), different weights could be applied \cite{mitchellFullyAdaptiveRadar2018}. Table \ref{table:Para_goals} summarizes typical choices of these parameters under specific goals.
\begin{table}[h!]
  \centering
  \caption{Parameters for different goals}
  \label{table:Para_goals}
  \begin{tabular}{|>{\centering\arraybackslash}m{4cm}|>{\centering\arraybackslash}m{2cm}|>{\centering\arraybackslash}m{2cm}|>{\centering\arraybackslash}m{2cm}|}
  \hline
  \textbf{Goal} & \textbf{$w_1$} & \textbf{$w_2$} & \textbf{$w_3$} \\ \hline
  Search for new target & Large & Small & Large \\ \hline
  LPI & Free & Large & Large \\ \hline
  Classification & Large & Small & Free \\ \hline
  Target tracking & Large & Large & Small \\ \hline
  Spectrum sharing & Small & Small & Large  \\ \hline
  \end{tabular}
  \end{table}
  In this study, we consider a CFMR with a 2-layer hierarchy that can switch between 3 tasks. Task 1 is for target searching, with corresponding weights set to $w_1=5$, $w_2=1$, $w_3=5$; Task 2 is for LPI, with corresponding weights set to $w_1=1$, $w_2=5$, $w_3=5$; Task 3 is for Spectrum sharing, with corresponding weights set to $w_1=1$, $w_2=1$, $w_3=5$. For all the tasks, the radar interacts with the environment with random actions to collect 10,000 trajectories to estimate the transition matrix $P$, and the value iteration algorithm is applied to find a near-optimal policy.
  
  \subsection{Baseline Methods and Evaluation Metrics}
  Because the proposed method can not only infer the reward function but also cluster the trajectories, we compare our method against several baseline approaches on both clustering and inference performance. The baseline methods we consider are:
  \begin{itemize}
      \item \textbf{MaxEnt IRL}: This method, introduced by Ziebart et al.\cite{ziebart2008maximum}, models the reward function with a linear model. It serves as a foundational approach for IRL problems.
      \item \textbf{Deep MaxEnt IRL\cite{wulfmeier2015maximum}}: An extension of MaxEnt IRL that uses deep neural networks to approximate the reward function. This method has been applied to estimate rewards of CR in \cite{zhu2022intelligent}. 
      \item \textbf{MLIRL}: This method uses multiple linear functions with EM to estimate and cluster trajectories. 
      \item \textbf{Gaussian Mixture Model (GMM) Clustering}: This method clusters the dataset based solely on its feature representations without considering the reward structure.
      \item \textbf{K-Means Clustering}: A popular unsupervised learning algorithm that partitions a dataset into $K$ clusters based on feature similarity.
      \item \textbf{RNN}: A widely used neural network architecture for sequence modeling. This model has been used in \cite{xuMethodFunctionalState2021,apfeldModellingLearningPrediction2019,liWorkModesRecognition2020} for radar work modes (resource management) recognition, and used in \cite{apfeldModellingLearningPrediction2019} for complex radar emitter behavior prediction.
  \end{itemize}
  To evaluate the performance, we use the following metrics:
  \begin{itemize}

    \item \textbf{Adjusted Rand Index (ARI):} The Adjusted Rand Index measures the agreement between two clustering assignments by considering all pairs of samples. Given true labels \( \{z_1, z_2, \ldots, z_n\} \) and predicted labels \( \{\hat{z}_1, \hat{z}_2, \ldots, \hat{z}_n\} \), the ARI is defined as:
    \[
    \text{ARI} = \frac{\sum_{i < j} \left[\mathbb{I}\left(z_i = z_j \right) \mathbb{I}\left(\hat{z}_i = \hat{z}_j \right) + \mathbb{I}\left(z_i \neq z_j \right) \mathbb{I}\left(\hat{z}_i \neq \hat{z}_j \right)\right] - \mathbb{E}[\text{RI}]}{\frac{1}{2} \left( n(n-1) \right) - \mathbb{E}[\text{RI}]}
    \]
    where \( \mathbb{I}(\cdot) \) is the indicator function, and \( \mathbb{E}[\text{RI}] \) is the expected value of the Rand Index under random labeling.

    \item \textbf{Normalized Mutual Information (NMI):} The Normalized Mutual Information quantifies the similarity between the clustering assignments of true labels \( \{z_1, z_2, \ldots, z_n\} \) and predicted labels \( \{\hat{z}_1, \hat{z}_2, \ldots, \hat{z}_n\} \). It is defined as follows:
    \[
    \text{NMI}(Z, \hat{Z}) = \frac{2 \cdot I(Z; \hat{Z})}{H(Z) + H(\hat{Z})}
    \]
    where \( Z = \{z_1, z_2, \ldots, z_n\} \), \( \hat{Z} = \{\hat{z}_1, \hat{z}_2, \ldots, \hat{z}_n\} \), \( I(Z; \hat{Z}) \) is the mutual information between \( Z \) and \( \hat{Z} \), and \( H(Z) \) and \( H(\hat{Z}) \) are the entropies of the true and predicted labels, respectively.
    \item \textbf{Action Prediction Error(APE)}: This metric measures the deviation of the actions predicted by the inferred policy from the actions taken in the true trajectories. It is defined as follows:
    \[
    \text{Action Prediction Error} = \frac{1}{N_{sa}} \sum_{\tau_i \in D }\sum_{a_i \in \tau_i } \mathbb{I}(a_i \neq \hat{a}_i)
    \]
    where \( N_{sa} \) is the total number of state-action pairs in the trajectory dataset \( D \), \( a_i \) is the true action taken at the \( i \)-th step, and \( \hat{a}_i \) is the action predicted by the IRL algorithm.

    \item \textbf{Expected Value Difference (EVD)}: This metric measures the difference in expected cumulative rewards between the expert policy \( \pi^* \) and the policy derived from the inferred reward function \( \pi_{\hat{R}} \).
    \[
    \text{EVD} = \left| \mathbb{E}\left[ \sum_{t=0}^{T} \gamma^t R(s_t, \pi^*(s_t),s_{t+1}) \right] - \mathbb{E}\left[ \sum_{t=0}^{T} \gamma^t R(s_t, \pi_{\hat{R}}(s_t),s_{t+1}) \right] \right|
    \]
    where \( \gamma \), \( T \), \( s_t \) are the same as defined before.
\end{itemize}
 \subsection{Results on single Intention IRL}
 We first compare the performance of different methods on the dataset produced by a single reward function. This time, only Task 1 is utilized. We use the near-optimal policy to generate trajectories under different circumstances to test the IRL algorithms.

\begin{figure}[htbp]
  \centering
  \begin{subfigure}{0.45\textwidth}
    \centering
    \includegraphics[width=\linewidth]{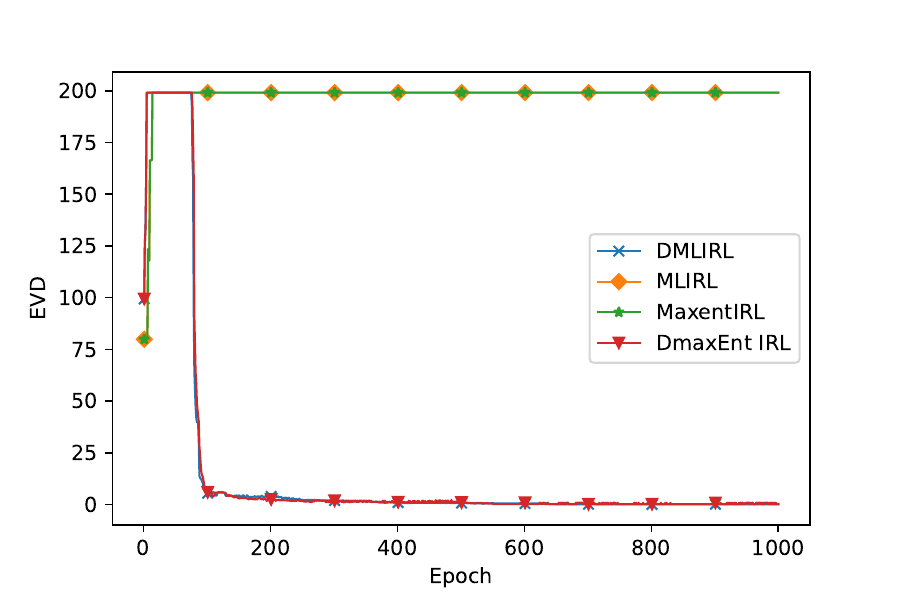}
  \end{subfigure}
  \hfill
  \begin{subfigure}{0.45\textwidth}
    \centering
    \includegraphics[width=\linewidth]{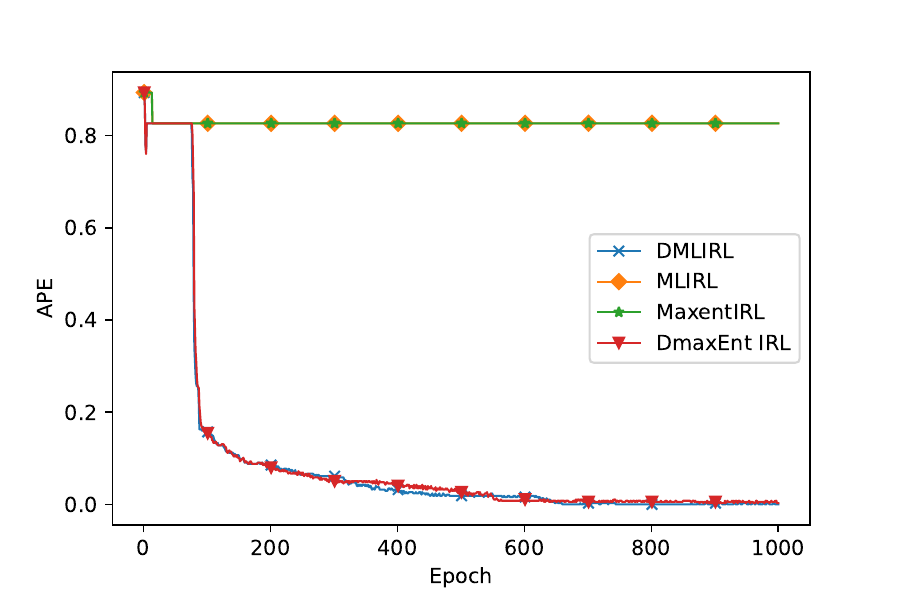}
  \end{subfigure}
  \caption{EVD and APE of different methods at a varying number of training iterations.}
  \label{fig:training_curve}
\end{figure}

First, we conduct the experiment in an ideal situation, where the ESM can correctly extract all the actions of the radar. 1000 demonstration trajectories are used, and the training curves are presented in Fig. \ref{fig:training_curve}.
 As observed, the model employing a deep neural network shows superior performance, quickly reducing the EVD to a low value. In contrast, the algorithm with linear weights completely failed to converge. This indicates that a neural network is necessary when inferring such complex reward structures. DMLIRL and DmaxEnt IRL, both method have nearly the same convergence speed and can completely replicate the behavior of the expert.

\begin{figure}[htbp]
  \centering
  \includegraphics{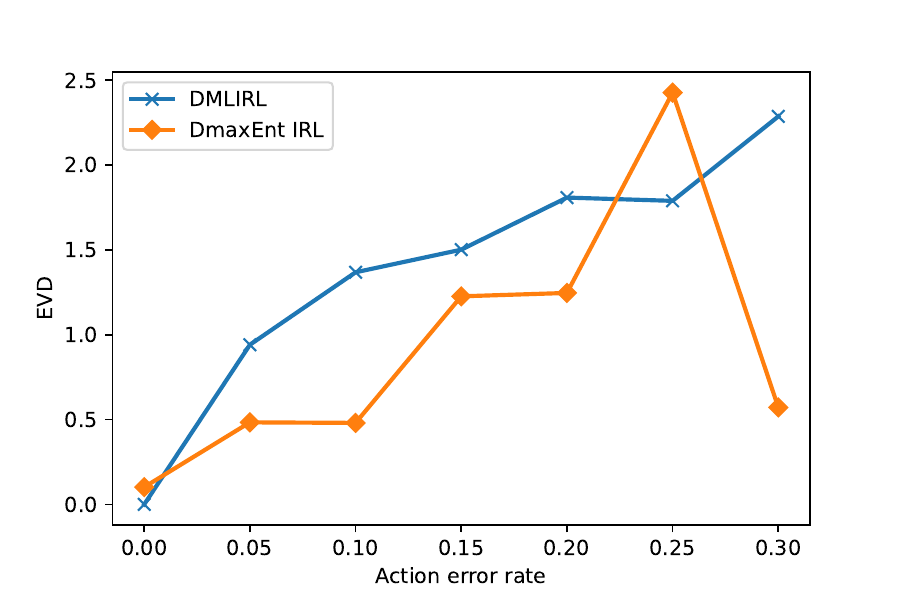}
  \caption{EVD of different methods at different action prediction error rates.}
  \label{fig:AER_EVD}
\end{figure}

Next, we test the performance under different ratios of action extraction errors. We only test for DMLIRL and DmaxEnt IRL as other methods failed to converge. It is assumed that each action is equally likely to be flipped to a wrong one due to non-ideal receiving of the ESM. We use the Action Error Ratio (AER) to represent the probability that any given action is misinterpreted. As observed in Fig. \ref{fig:AER_EVD}, the EVD only slightly increased as the Action Error Ratio increases, up to 0.3. This indicates that both DMLIRL and DmaxEnt IRL is not sensitive to noisy observations. DmaxEnt IRL slightly outperforms DMLIRL as maximizing the entropy makes minimal commitments and is the least wrong\cite{aroraSurveyInverseReinforcement2021}.

 Lastly, we limited the number of available trajectories to investigate how different IRL algorithms perform with varying amounts of training data. The number of training trajectories ranged from 10 to 1000, and we measured both the EVD and the Action Prediction Error (APE) to assess the effectiveness of each algorithm.

\begin{figure}[H]
  \centering
  \begin{subfigure}{0.45\textwidth}
    \centering
    \includegraphics[width=\linewidth]{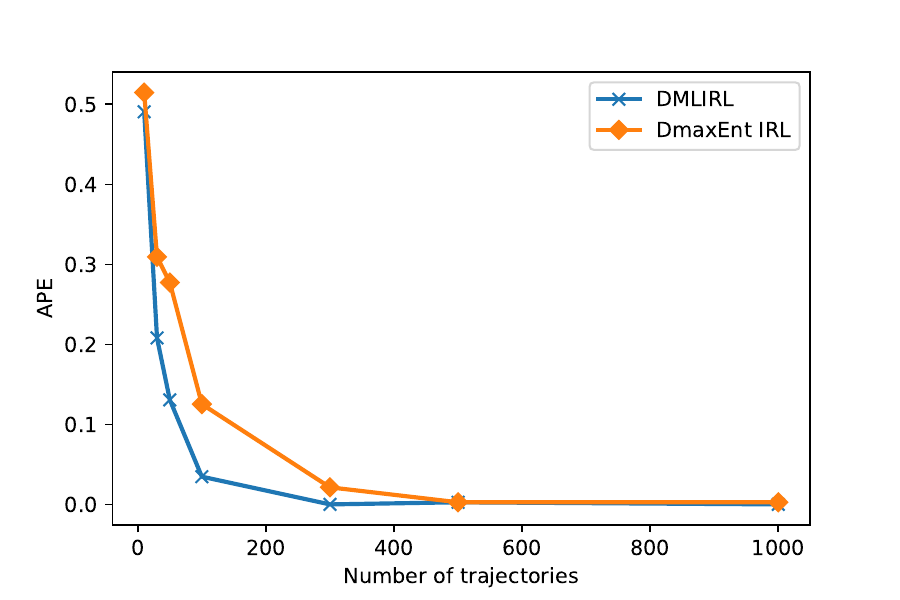}
  \end{subfigure}
  \hfill
  \begin{subfigure}{0.45\textwidth}
    \centering
    \includegraphics[width=\linewidth]{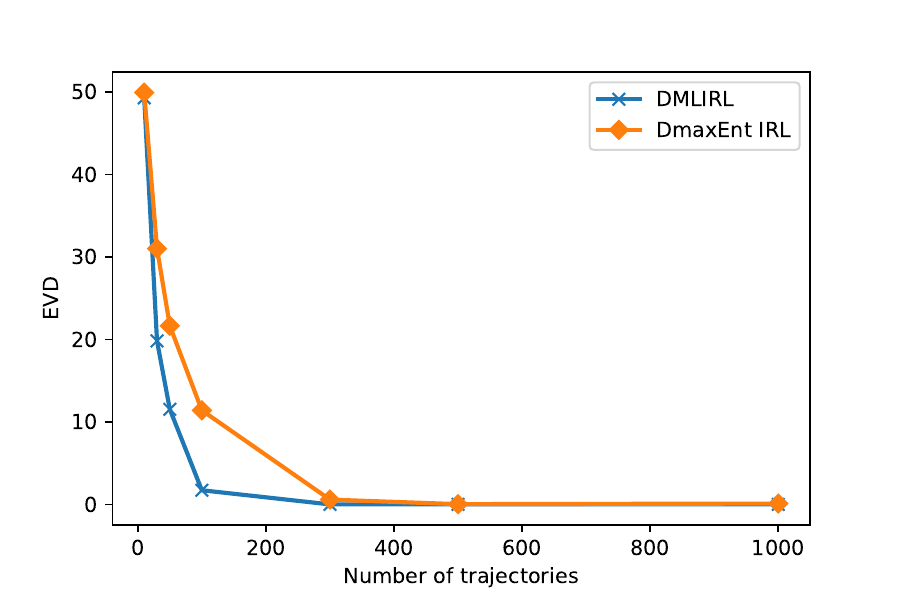}
  \end{subfigure}
  \caption{EVD and APE of different methods under different number of training trajectories.}
  \label{fig:N_traj}
\end{figure}

The results are presented in Fig. \ref{fig:N_traj}. Both EVD and APE decrease sharply as the number of trajectories increases. This is because both the estimation of the transition matrix and the diversity of the expert demonstrations can affect the performance. The more training trajectories available, the more accurate the transition matrix estimation becomes, and the more diverse the expert demonstrations are, leading to better generalization and reduced error in policy prediction. Noticeably, when the dataset contains fewer than 300 trajectories, the  Maximum-Likelihood-based method outperforms the Maximum-Entropy-based method. This is because in ML-based IRL, the estimated transition matrix only affect the training through value iteration, whereas for Maxent-based IRL, both the value iteration and the policy propagation (see algorithm \ref{alg:maxent}) can be affected.   

 \subsection{Results on multiple Intentions IRL}
\begin{figure}[H]
  \centering
  \begin{subfigure}{0.45\textwidth}
    \centering
    \includegraphics[width=\linewidth]{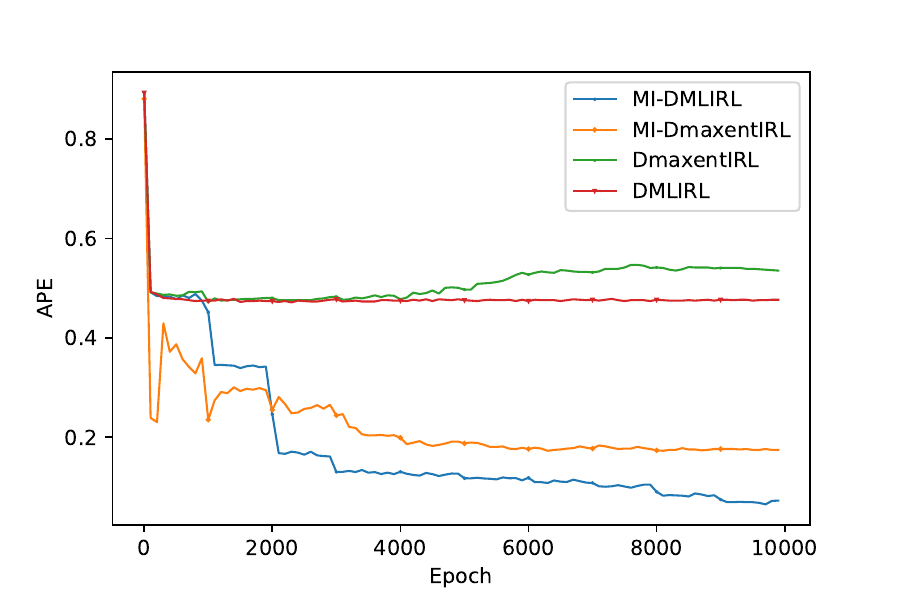}
  \end{subfigure}
  \hfill
  \begin{subfigure}{0.45\textwidth}
    \centering
    \includegraphics[width=\linewidth]{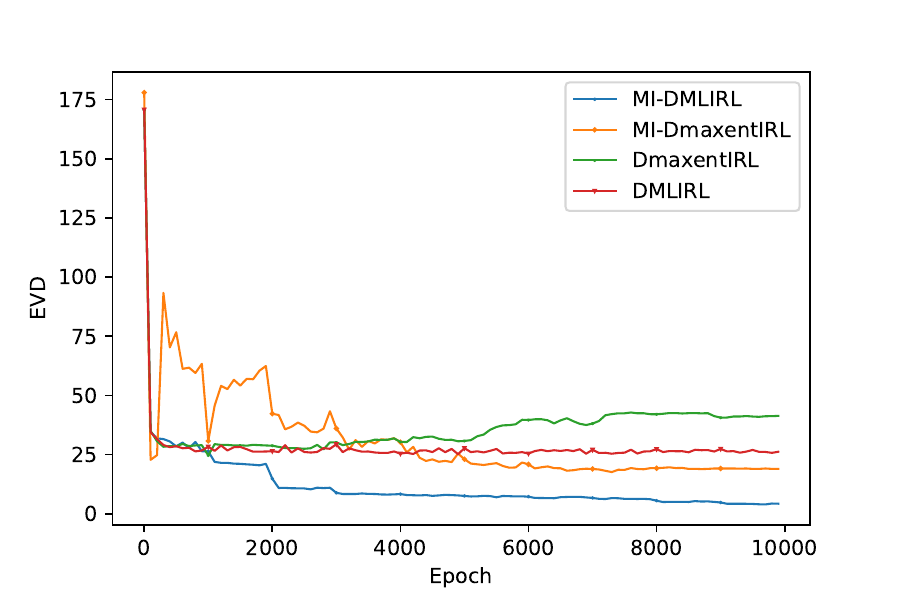}
  \end{subfigure}
  \caption{Training curves for different IRL method on the dataset containing task 1,2 and 3.}
  \label{fig:training_curve_mi}
\end{figure}

 In this section, we assume the radar can perform different tasks in different trajectories. Thus, the dataset is generated by Task 1, 2, and 3, and the reconnaissance side has to not only estimate the reward functions but also cluster the trajectories. 1000 demonstration trajectories for each task are generated and mixed to evaluate the performance. The training curves for different IRL methods are presented in Fig. \ref{fig:training_curve_mi}, and final testing results are summarized in Table \ref{tab:performance_comparison}.

 \begin{table}[H]
  \centering
  \caption{Performance Comparison of Different Methods}
  \label{tab:performance_comparison}
  \begin{tabular}{lcccc}
  \hline
  \textbf{Method} & \textbf{NMI} & \textbf{ARI} & \textbf{EVD} & \textbf{APE} \\
  \hline
  DMLIRL & - & - & 26.21 & 0.48 \\
  DmaxEnt IRL & - & - & 41.36 & 0.54 \\
  MI-DMLIRL & 1.00 & 1.00 & 4.26 & 0.07 \\
  MI-DmaxEnt & 0.73 & 0.57 & 18.97 & 0.17 \\
  RNN & 1.00 & 1.00 & - & 0.14 \\
  K-means & 0.57 & 0.47 & - & - \\
  GMM & 0.63 & 0.54 & - & - \\
  \hline
  \end{tabular}
  \end{table}
  For reward estimation, the methods with just a single model failed to converge as the EVD is significantly higher than those of multiple intentional IRL models. This is because different trajectories of the CMFR can come from completely different tasks, which can not be explained with a reward model. Compared with the Maxent-based IRL, ML-based IRL better suits EM based the training. As shown in Fig. \ref{fig:training_curve_mi}, for MI-DMIRL both EVD and APE decreased significantly for each EM iteration and quickly converge to near optimal values. However, the training for MI-DmaxentIRL is not so stable, and it finally stuck in a local minimal.

  For clustering, the MI-DMLIRL method demonstrates perfect clustering with both NMI and ARI values reach 1, which is the same as the supervised-trained RNN. In contrast, MI-DmaxEnt fails to make perfect alignment with the ground truth task labels as its learning stuck at local minimal. However, it still performances better than traditional clustering methods such as K-means and GMM because they only use the action information.

  For actions prediction, the MI-DMLIRL method excels in predicting the correct actions, as evidenced by its low APE of 0.07. In contrast, MI-DmaxEnt shows a higher APE of 0.17 as its convergence to suboptimal solutions impairs its ability to consistently predict actions across all tasks. Interestingly, the RNN model, although not using the state-action relations, also achieves a low APE of 0.14. This again proofs that context information is important for complex radar emitter behavior prediction\cite{apfeldValueMemoryMarkov2021}.
 \section{Conclusion}
 The paper introduces a deep multi-intentional inverse reinforcement learning method for effectively understanding and analyzing cognitive multi-function radars. By leveraging an EM algorithm with neural networks, the proposed method successfully clusters trajectories and estimates multiple reward functions in the presence of complex and multi-task radar settings. The simulation results showcase the robustness and superior performance of DMLIRL+EM in comparison to single expert IRL methods and traditional pattern recognition approaches.
\bibliographystyle{ieeetr} 
\bibliography{manuscript.bib} 
\end{document}